\documentclass[prl,twocolumn,english,superscriptaddress,floatfix]{revtex4}
\usepackage{graphicx}
\usepackage{bm, amsmath, amssymb}
\usepackage{color}
 \usepackage{soul}
\usepackage{pdfpages}
\usepackage[T1]{fontenc}
\usepackage[latin9]{inputenc}
\usepackage{amstext}
\usepackage{bbold}
\usepackage{esint}

\def\be{\begin{equation}}
\def\ee{\end{equation}}
\def\bea{\begin{eqnarray}}
\def\eea{\end{eqnarray}}

\usepackage{babel}

\begin{document}

\title{ Pre- and post-selected averages and correlation functions of a continuously monitored qubit.}
\title{ Pre-selection, Post-selection, and Correlation Functions of a Continuously Monitored Superconducting Qubit}
\title{Correlations of the time dependent signal and the state of a continuously monitored quantum system}

\author{N. Foroozani}
\affiliation{Department of Physics, Washington University, St.\ Louis, Missouri 63130}
\author{M. Naghiloo}
\affiliation{Department of Physics, Washington University, St.\ Louis, Missouri 63130}
\author{D. Tan}
\affiliation{Department of Physics, Washington University, St.\ Louis, Missouri 63130}
\author{K. M\o lmer}
\affiliation{Department of Physics and Astronomy, Aarhus University, Ny Munkegade 120, DK-8000 Aarhus C, Denmark}
\author{K. W. Murch*}
\affiliation{Department of Physics, Washington University, St.\ Louis, Missouri 63130}
\date{\today}

\begin{abstract}

In quantum physics, measurements give random results and yield a corresponding random back action on the state of the system subject to measurement. If a quantum system is probed continuously over time, its state evolves along a stochastic quantum trajectory. To investigate the characteristic properties of such dynamics, we perform weak continuous measurements on a superconducting qubit that is driven to undergo Rabi oscillations. From the data we observe a number of striking temporal correlations within the time dependent signals and the quantum trajectories of the qubit, and we discuss their explanation in terms of quantum measurement and photodetection theory.

\end{abstract}

\maketitle


A collection of quantum oscillators prepared in the same initial state will oscillate in phase until decoherence mechanisms cause the ensemble to dephase. If the oscillators are weakly measured, the resulting average signal will reflect the ensemble behavior of the oscillators and exhibit  damped oscillatory behavior.  In addition, if the quantum oscillator is not in an eigenstate of the measured observable, additional dephasing will be present as mandated by the Heisenberg uncertainty principle.  Thus, stronger measurements result in faster damping of the observed oscillatory signal. But, what average signal does one expect to observe if instead a projective measurement is used to post-select \cite{ahar10} the oscillators in the same final state? In this Letter, we show with experimental data that the average post-selected signal is exactly the time reverse of the pre-selected one where the oscillators are prepared in the same initial state. This temporal correlation between qubit state populations and observed measurement signals  is an example of the interwoven nature of measurement signals and quantum trajectories that have become available with current experiments on superconducting qubits \cite{murc13traj,webe14,camp13}. Inspired by their use in the analysis of stochastic processes, we address different two-time correlation functions, available from the experiments and discuss how their properties may be analyzed by quantum measurement theory and by master equation approaches.


In quantum measurements, the detection of light conveys information about the state of the emitter and induces corresponding backaction on its quantum state. Temporal correlations of light reveal purely quantum effects and have been essential to the development of quantum optics \cite{hanb56,glau63} which traditionally applies the master equation and the quantum regression theorem for evaluating the observed measurement signal. Recent experiments have harnessed the quantum trajectory description \cite{carm93,murc13traj,webe14,tan15} to track the emitter's dynamical evolution. In this work, we take such analysis one step further and investigate two-time correlation functions between the measurement signal and the quantum state. As expected from quantum optics, the correlations associated with quantum measurements at different times can reveal purely quantum effects for example by testing the assumptions of macrorealism via the Leggett-Garg inequality \cite{legg85,rusk06,will08,pala10,groe13,whit15}. 

We perform continuous weak measurement on a superconducting qubit in the energy basis while the qubit is resonantly driven to produce Rabi oscillations \cite{vija12,webe14,camp13}.  Measurements are executed in the dispersive regime of cavity quantum electrodynamics \cite{bois09} where the interaction between the qubit and a mode of the cavity is given by the interaction Hamiltonian, $H_\mathrm{int.} = -\hbar \chi a^\dagger a \sigma_z$, where $a^\dagger (a)$ is the creation (annihilation) operator for the cavity mode, $\sigma_z$ is the Pauli pseudo-spin operator that acts on the qubit in the energy basis and $\chi$ is the dispersive coupling rate. Homodyne probing of the cavity can thus be used to conduct both weak and strong measurements of the qubit state in the $\sigma_z$ basis.  We describe these measurements by the theory of POVMs (positive operator valued measures) which relates the homodyne voltage signal $V$ to the qubit state by the operators \cite{jaco06},
\begin{eqnarray}
\Omega_V = (2 \pi a^2 )^ {-1/4}e^{-(V - \sigma_z)^2/4a^2}.
\end{eqnarray}
The probability of detecting a homodyne voltage $V$, at time $t$, $P(V_t) = \mathrm{Tr}(\Omega_V \rho_t \Omega_V^\dagger)$, depends on the   density matrix $\rho_t$, and it is the sum of two Gaussians centered at $V=\pm 1$ and weighted by the qubit state populations $\rho_t^{00}$ and $\rho_t^{11}$. Depending on the variance $a^2 = 1/(4 k \eta \Delta t)$, where $k$ represents the measurement strength, $\eta$ is the quantum measurement efficiency and $\Delta t$ is the integration time of the measurement, this operator describes both strong (projective) and weak measurements of the qubit.

If the measurement is strong, the variance $a^2$ is small and the measurement outcome unambiguously belongs to one of the two disjoint Gaussian distributions \cite{johnson12}. If, conversely, the measurement is weak, the variance $a^2$ is large, and $P(V)$ can be approximated by a single Gaussian distribution centered at the expectation value of $\sigma_{z}$,
\begin{equation} \label{weakP}
P(V) \simeq (2 \pi a^2 )^ {-1/2}e^{-(V -\langle \sigma_z \rangle)^2/2a^2}.
\end{equation}
In this regime, $V$ thus provides a noisy estimate of $\langle \sigma_z \rangle$.

In addition to the assignment of outcome probabilities, the POVM operators describe the effect of the measurement on the quantum state, $\rho\rightarrow \Omega_V \rho \Omega_V^\dagger$. In the limit of strong measurements, a result $V\simeq \pm 1$ corresponds to the projection operators $\Omega_{\pm z} = (\sigma_z\pm1)/2$ on the $\pm z$ qubit eigenstates.  For weak measurement, the backaction on the state is small and the evolution of the density matrix associated with a Hamiltonian $H_R$ (describing unitary evolution) and a measurement signal $V(t)$ can be obtained from the stochastic master equation \cite{jaco06,gamm13,tan15},
\begin{eqnarray} \label{SME}
\dot{\rho}  = \frac{1}{i\hbar}[H_R,\rho] + k (\sigma_z \rho \sigma_z  - \rho)\nonumber \\
 + 2 \eta k (\sigma_z \rho + \rho \sigma_z - 2 \mathrm{Tr}(\sigma_z \rho)\rho) V(t).
\end{eqnarray}
If we disregard the measurement outcome, or average over its values, the qubit  density matrix is still subject to dephasing due to the probing and  solves a deterministic master equation,
\begin{equation} \label{ME}
\dot{\rho}\vert_{det} =\frac{1}{i\hbar}[H_R,\rho] + k (\sigma_z \rho \sigma_z  - \rho).
\end{equation}

In our experiment, the qubit is formed from the two lowest energy levels of a transmon circuit \cite{koch07} (see Fig. 1(a)). The transmon is dispersively coupled to a three dimensional waveguide cavity \cite{paik113D}. The qubit transition frequency  is $\omega_q/2 \pi = 4.0033$ GHz and the cavity transition frequency is 6.9914 GHz with bandwidth $\kappa/2\pi = 9.88$ MHz and dispersive coupling rate $\chi/2\pi = -0.425$ MHz. Since $2|\chi|\ll \kappa$, qubit state information is encoded in a single quadrature of a transmitted microwave tone that probes the cavity. Qubit decay from the excited to the ground state is slow and can be ignored for the time scales of our experiments. The reflected signal is amplified by a near-quantum-limited Josephson parametric amplifier \cite{hatr11para} and demodulated and digitized after further amplification at room temperature. Further details of the experimental setup can be found in supplemental information \cite{supp}.

We prepare the qubit in an initial state by applying a  $\pi/2$-rotation about the $y$ axis and then making a projective measurement $\Omega_{\pm z}$.  This projective measurement can be used to herald either $+z$ or $-z$  as the initial state or, if the result of the projective measurement is ignored, to prepare an initial mixed state. Following this preparation, the qubit is subject to continuous rotation given by $H_\mathrm{R} = \hbar\Omega_R \sigma_y/2$, where $\Omega_R/2\pi =1.16$ MHz is the Rabi frequency, and probing as given by the operators $\Omega_V$. The experiment is repeated several times to form an ensemble of measurement signals $\{V_i(t)\}$ which is analyzed in a post processing step.


Fig.\ 1(b) displays how we obtain the measurement signals and how the qubit is initially heralded in the $+z$ state by a strong measurement. Using the stochastic master equation (\ref{SME}) we propagate the density matrix according to the individual, weak measurement signals. The average voltage signal $\overline{V(t)}$ in Fig.1(c) shows damped Rabi oscillations corresponding to the gradual dephasing of the qubit due to the measurement interaction. In Fig. 1(d), the qubit expectation value, $z_i(t) = \mathrm{Tr}(\rho_i(t) \sigma_z)$ calculated by solution of (\ref{SME}), is shown for several of the individual quantum trajectories \cite{wisebook,murc13traj}, confirming that the damping of the measurement signal is due to dephasing of the ensemble.

 \begin{figure}\begin{center}
 \includegraphics[angle = 0, width = .47\textwidth]{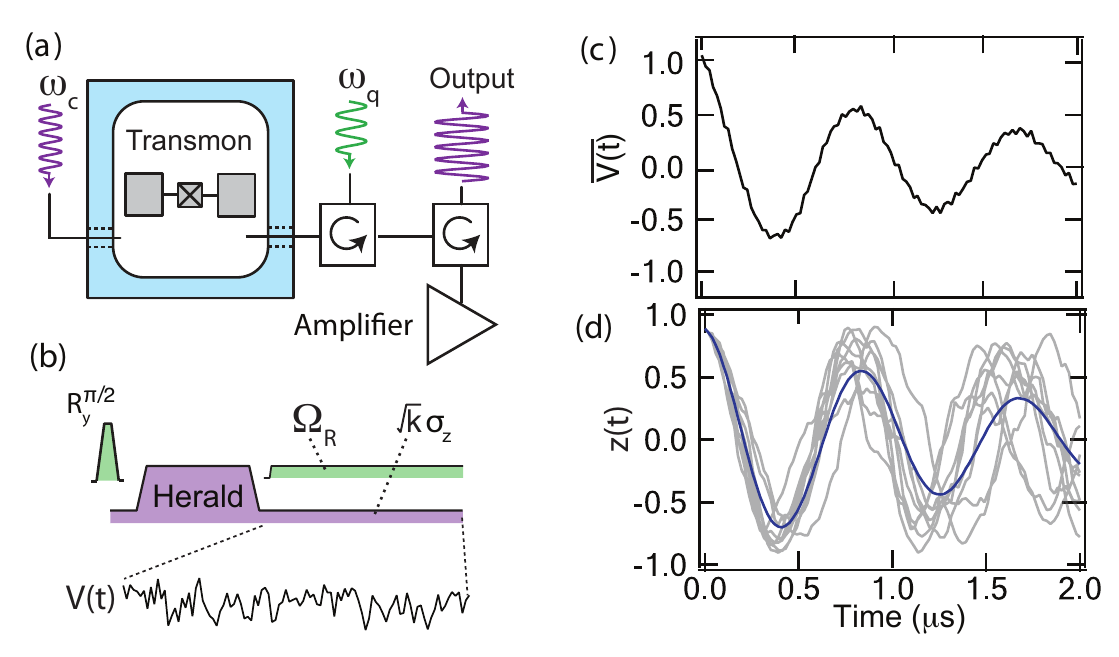}
\end{center}
\caption{\small{ (a) Experimental setup consisting of a transmon circuit coupled to a 3D cavity. (b) Continuous probing of a qubit driven with a Rabi frequency $\Omega_R/2\pi=1.16\ \textrm{MHz}$ after its heralded preparation in $+z$ or $-z$. The probe is given by the measurement operator $\sqrt {k} \sigma_z$, where $k = 4\chi^2\bar{n} /\kappa$, parametrizes the measurement strength $(k/2 \pi =$ 95 kHz) and $ \eta = 0.35$ is the quantum measurement efficiency. (c) Pre-selected average of the output measurement signals, conditioned on the heralded preparation of $+z$. (d) The expectation value $z_i(t)$ in a number of individual quantum trajectories (gray lines) and their mean value (blue line), which is in agreement with the mean signal shown in the upper panel. }}
\end{figure}



In Fig. 2(a), the black curve displays the average of the measurement signals conditioned on the outcome of a final projective measurement which post-selects the qubit in the $+z$ state.  The resulting average exhibits an oscillatory signal that is damped \emph{backwards} in time \cite{camp13}.  Surprisingly, this is exactly the time reverse of the damped signal observed in Fig.\ 1(c) and it exhibits the same full contrast at the final time $t=T$ that we observe for the pre-selected average at $t=0$.  

Fig.\ 2(b) displays a sample of the trajectories $z_i(t)$, which are propagated forward from the initial state using the stochastic master equation. Since the post-selection average is conditioned only on the final measurement, a roughly equal number of the trajectories, included in this average, are initially detected in the $+z$ and $-z$ states. As shown in Fig.\ 2(c) immediately prior to the post-selection, the trajectories take on several different values of $z_i(T)$ between $\pm 1$ , and thus it may be surprising that the average measurement signal (black curve in Fig.\ 2(a)) exhibits full contrast oscillations at the end of the sequence.

\begin{figure}\begin{center}
\includegraphics[angle = 0, width = .45\textwidth]{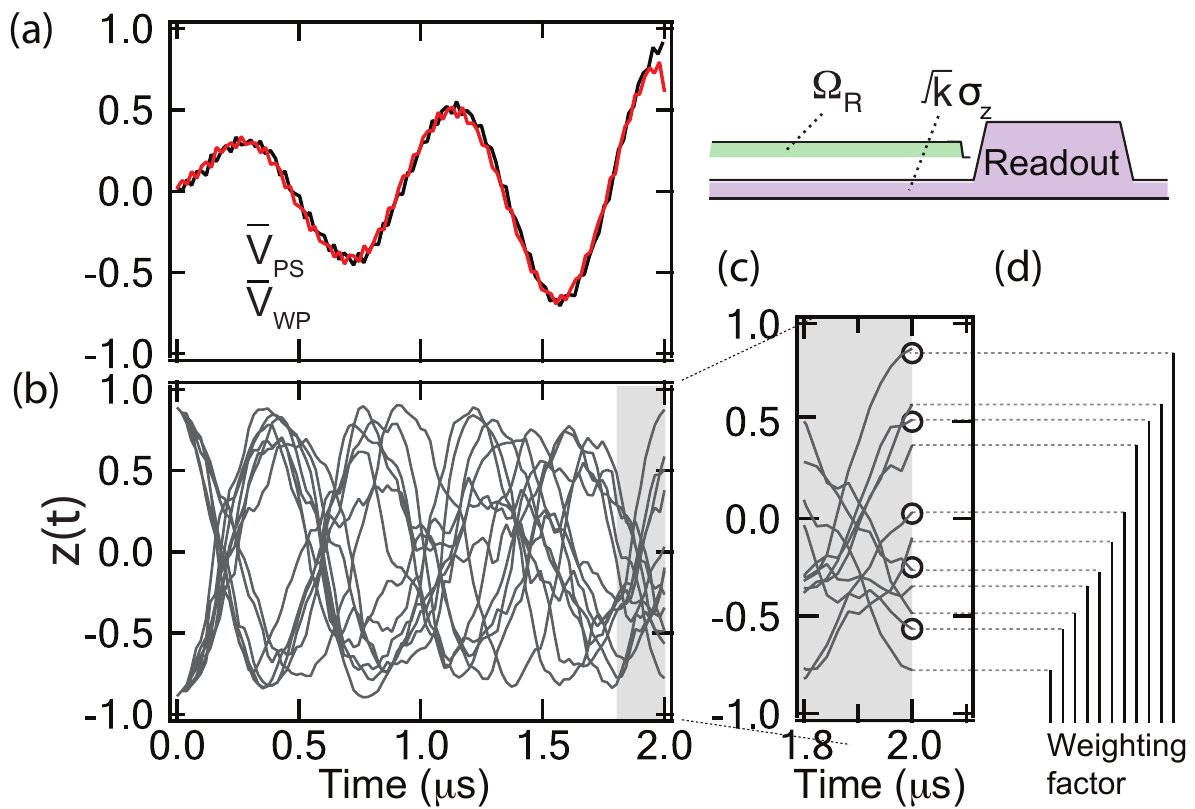}
\end{center}
\caption{\small{(a) The average of measurement signals conditioned the post-selection of $+z$ (black trace).  The red trace is the average of \emph{all} of the measurement signals weighted by the final population $\rho_i^{00}(T)$. (b) The quantum trajectories whose measurement outcomes contribute to the post-selected and weighted averages in (a) are shown as $z_i(t)$ evolves forward in time from the initial state.   (c) The magnified view of the last time segment shows that the trajectories populate the qubit eigenstates very differently. Trajectories that are post-selected in the $+z$ state are indicated with the open circles. The measurement signals $V_i$ corresponding to this sub-ensemble form the average $\overline{V}_\mathrm{PS}$.  (d) The vertical bars indicate the relative weighting applied to all of the measurement signals that contribute to $\overline{V}_\mathrm{WP}$. }}
\end{figure}

It is tempting to assume that the time reversed damped signal is merely a quirky consequence of post-selection \cite{ahar10,vaid13}, but we will now show that the same behavior can be obtained by a suitably weighted average over all trajectories without post-selection.  Consider the stochastic master equation (\ref{SME}) yielding for each experimental run an independent realization $\rho_i(t)$ of the time dependent conditioned density matrix of the qubit. We now average all the measurement signals $V_i(t)$, weighted according to the final state probability $P(i,+z) = \rho_i^{00}(T)$,
\begin{equation}
\overline{V}_{\mathrm{WP}} =  \frac{\sum_i \rho_i^{00}(T) V_i (t)}{\sum_i \rho_i^{00}(T)}. \label{weighted}
\end{equation}
Fig.\ 2(d) indicates using vertical bars the relative contribution of each of the signals to the average. As shown by the red curve in Fig.\ 2(a), this weighted average is in excellent agreement with the post-selected average.

In order to explain the equivalence between the post-selected and the weighted average we observe that the post-selected mean value can be written as,
\begin{equation}
\overline{V}_\mathrm{PS}= \frac{\sum_i n_i V_i (t)}{ \sum_i n_i},  \label{post}
\end{equation}
where the stochastic variable $n_i = 0, 1$ depends on the post-selection result for the $i^{th}$ element of the ensemble.  Since the probability of the post-selection measurement yielding $+z$ is $\rho _i^{00}(T)$, in the limit of a large ensemble, $\sum_i n_i V_i (t)$ approaches $\sum_i \rho_i^{00}(T)V_i (t)$. Similarly, the denominator $\sum_i n_i$ approaches $\sum_i \rho_i^{00}(T)$, and $\overline{V}_{WP}=\overline{V}_{PS}$  because the individual trajectories faithfully predict the probability of the final projective state measurement.

To explain the symmetry of the pre- and post-selected averages, let us consider the joint probability $P(V_t,\pm z_T)$  for the measurement outcome $V_t$ at time $t$ and a projective measurement on $\pm z_T$ at time $T$. This probability factors as the product of the probability for the first measurement outcome $P(V_t)$ and the conditional probability $P(\pm z_T\vert V_t)$,
\begin{eqnarray}\label{probabilityL}
P(V_t, \pm z_T) = P(V_t) \cdotp \mathrm{Tr} \{  e^{L(T-t)}[\frac{\Omega_V \rho\Omega_V^{\dagger}}{P(V_t)}] |\pm z\rangle \langle \pm z|\},
\end{eqnarray}
where $\rho$ is the state prior to the weak measurement of $V_t$, $\Omega_V \rho \Omega_V^\dagger/P(V_t)$ yields the normalized state conditioned on the outcome $V_t$, and $e^{L(T-t)}$ denotes the linear propagator from time $t$ to $T$ of the density matrix according to the master equation (\ref{ME}).

It is convenient to express the deterministic master equation evolution from time $t$ to $T$ as a Kraus map, $ e^{L(T-t)}[\rho] = \sum_\alpha K_\alpha \rho K_\alpha^\dagger$, with operators obeying $\sum_\alpha K_\alpha^\dagger K_\alpha = I$ \cite{krau, Heyashi03}.
By making use of the cyclic properties of the trace, we can  shift the Kraus operators $K_\alpha$ to the right hand side of the expression (\ref{probabilityL}), yielding,
\begin{multline} \label{joint}
P(V_t, \pm z_T) = \mathrm{Tr} \{ \Omega_V \rho(t) \Omega_V^{\dag} \sum_{\alpha}  K_{\alpha}^{\dag} |\pm z\rangle \langle \pm z|K_{\alpha}\}\\
=\mathrm{Tr} [ \Omega_V \rho(t) \Omega_V^{\dag}E(t)],
\end{multline}
where $E(t)$ is defined as the operator expression involving the final state $|\pm z\rangle \langle \pm z|$ and the adjoint Kraus map. We do not need the explicit form of the Kraus operators, as $E(t)$ can be found by solving the (adjoint) master equation,
\begin{equation} \label{EME}
\frac{dE}{d(-t)} =\frac{-1}{i\hbar}[H_R,E] + k (\sigma_z E \sigma_z  - E),
\end{equation}
backwards in time from the state $E(T)= |\pm z\rangle \langle \pm z|$ to $t$.

While it may seem unusual to discuss the probability and weighted average of a measurement outcome $V_t$ conditioned on a \emph{later} projective measurement, such analyses have recently been proposed \cite{gamm13} and successfully applied to experiments \cite{tan15, camp13, ryba15}.
%
If we do not condition on the final measurement and set $E(t)\propto I$ in Eq.(\ref{joint}), we recover the usual prediction, $\overline{V}=\langle \sigma_z\rangle = \rho^{00}(t)-\rho^{11}(t)$, while if we condition on the outcome of the last and not the initial $\sigma_z$ measurement, $\rho$ is proportional to the identity  matrix in (\ref{joint}), and $\overline{V}_\mathrm{PS} = E^{00}(t)-E^{11}(t)$. The only difference between the forward evolution (\ref{ME}) of $\rho(t)$ and the backward evolution (\ref{EME}) of $E(t)$ is the sense of rotation of the damped Rabi oscillations. They therefore yield identical heralded predictions (see analytical expressions in the supplementary information) for the voltage signal.

 \begin{figure}\begin{center}
 \includegraphics[angle = 0, width = .47\textwidth]{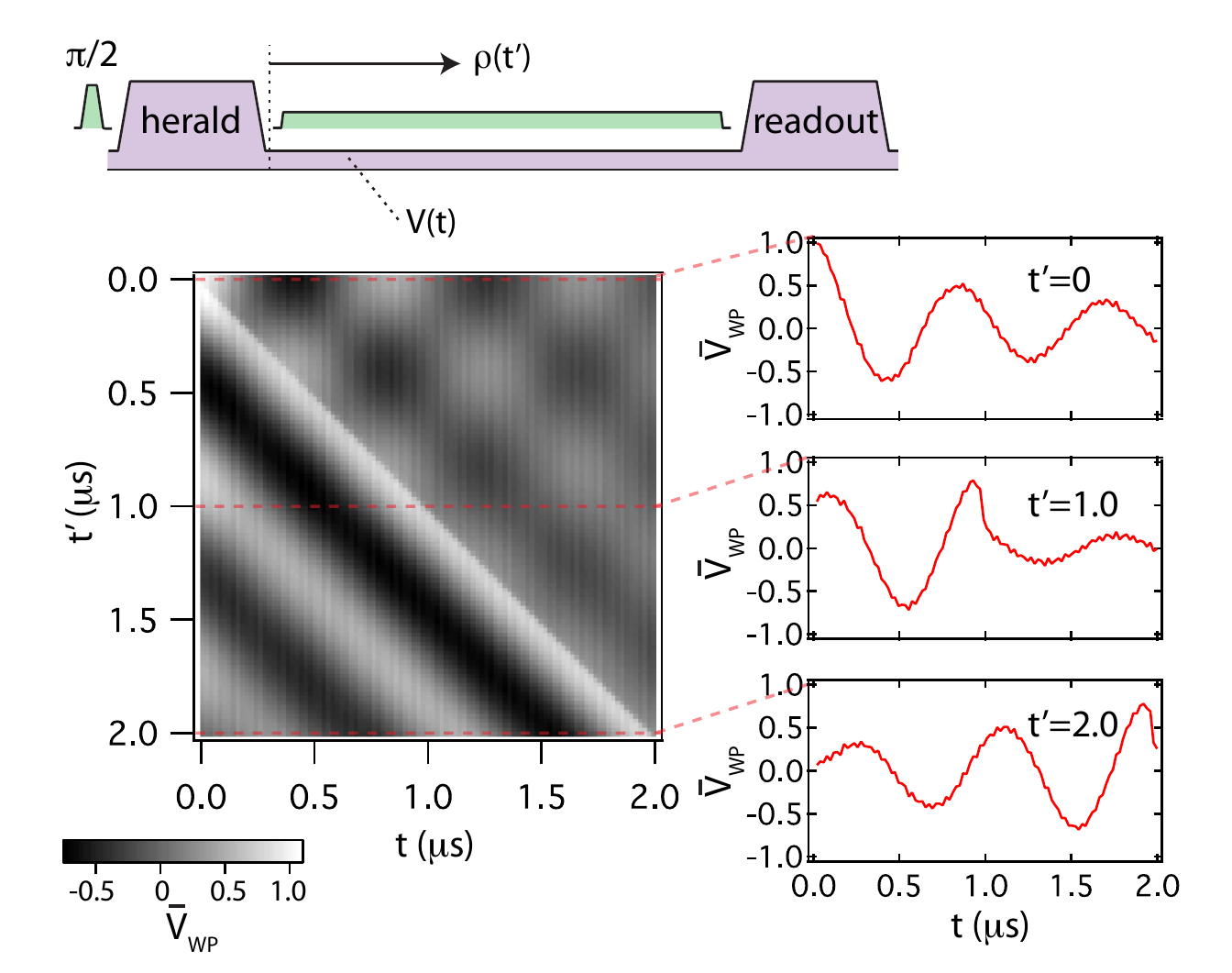}
\end{center}
\caption{Two-time correlation function between the measurement signal $V_i(t)$ and the inferred qubit density matrix element $\rho_i^{00}(t')$. The side panels show the correlations as function of $t$ for fixed values of $t'$.}
\end{figure}

While mysterious action from the future through post-selection evokes fascinating scientific debate \cite{ahar10,vaid13,ahar11,ahar11_2} the predictions we make for the post-selected or weighted averages merely reflect the correlation between the qubit observables at different times. Such correlations are central in the quantum optical characterization of light sources, and the time evolution between times $t$ and $T$ in our Eq.(\ref{joint}), indeed, appears in a very similar manner in the quantum regression theorem \cite{gard04book} when it is applied to calculate field intensity and amplitude correlations \cite{xu15}.

To examine further the relationship and the temporal correlations between the measurement signal and the inferred qubit density matrix, we show in Fig. 3 the "hybrid" correlation function between the measured signal and the inferred qubit state,
\begin{equation}
\overline{V}_{WP} (t,t')= \frac{\sum_i \rho_i^{00}(t') V_i (t)}{\sum_i \rho_i^{00}(t')}, \label{twotime}
\end{equation}
obtained as an average over all the experimental runs of the voltage signal $V_i(t)$ at time $t$ weighed by the qubit excited state population $\rho_i^{00}(t')$ at time $t'$ .

The side panels in figure 3 show $\overline{V}_{WP} (t,t')$ as a function of $t$ between $0$ and $T$ for fixed values of  $t'$.  We clearly observe a change in the behavior of the function at $t=t'$ which suggests a different correlation regime before and after $t'$. This has a natural explanation: $V_i(t)$ is the sum of a term proportional to $\langle \sigma_z \rangle$ and a white noise term $W(t)$.  When $t < t'$, the noise term $W(t)$ contributes a change of  the conditioned density matrix due to (\ref{SME}) hence it affects all later values of $\rho_i^{00}(t')$. We thus expect that until $t = t'$, the product in Eq.\ (\ref{twotime}) will contain a contribution that is quadratic in the white noise term $W(t)$, while for $t > t'$, $W(t)$ is uncorrelated with the earlier value $\rho_i^{00}(t')$, and hence it averages to zero in the sum over $i$ in Eq. (\ref{twotime}).

For $t < t'$, $\overline{V}_{WP} (t,t')$ can be calculated by the analysis that we applied to the post selected data (by setting $T=t'$) and it exhibits all the features we have explored so far. Even though we do not perform a projective measurement at time $t'$, due to the equivalence of $\overline{V}_\mathrm{PS}$ and $\overline{V}_\mathrm{WP}$, the function plotted in the region with $t < t'$ can be calculated from (\ref{joint}), with  $E(t)$ propagated backwards from the state $|+z\rangle \langle +z|$ at time $t'$ \cite{supp}.

We recall that $\rho_i^{00}(t')=(\langle \sigma_z (t') \rangle_i + 1)/2$, and $V(t') = \langle \sigma_z (t') \rangle + W(t')$, where $W(t')$ has zero mean and is uncorrelated with all previous quantities. For $t<t'$, this leads to an alternative expression for $\overline{V}_{WP} (t,t')$ involving only the mean value and the two-time correlation function of measured quantities,
\begin{equation} \label{G2}
\overline{V}_\mathrm{WP}(t,t')= \frac{\overline{V_i(t)V_i(t')}+\overline{V_i(t)}}{\overline{V(t')}+1}.
\end{equation}
Note that the signal-state and the signal-signal correlations being averaged  in the different expressions for $\overline{V}_\mathrm{WP}$ are not identical and equality between them hold only for large ensembles of measurement records.


It is interesting to observe that for $t < t'$ where the state-signal correlation function can be expressed solely in terms of experimental signal  correlations (\ref{G2}), the past quantum state formalism provides a deterministic theory for the correlation function (\ref{twotime}). This is, indeed, intimately related to the conventional use of the quantum regression for the same purpose in quantum optics  \cite{xu15}. For $t  > t'$, the correlation function may instead be written $\overline{V}_{WP} (t,t') = \frac{\sum_i \rho_i^{00}(t') (2\rho_i^{00} (t)-1)}{\sum_i \rho_i^{00}(t')}$. Such correlations and higher moments of the quantum trajectory solutions do not obey any simple deterministic equation, and for each possible density matrix $\rho(t')$ resulting from the preceding stochastic measurement record (\ref{SME}), we must propagate $\rho(t)$ by Eq.(\ref{SME}) or (\ref{ME}) to determine $\overline{\rho^{00}(t')\rho^{00}(t)}$.

 Two-time correlation functions like $\overline{\rho^{00}(t')\rho^{00}(t)}$ have so far not found applications in quantum optics. The recent progress in circuit QED, however, has made it possible to assess the dynamics of single quantum systems  and thus demands a more detailed characterization of the stochastic system dynamics during measurements. Recently, a theory was presented for the most likely path of a stochastically evolving density matrix \cite{chan13, webe14, chan15}, and correlation functions, indeed, present a natural quantitative characterization of such a theory. Finally, stochastic state dynamics is an unavoidable, and even useful, component of precision metrology and parameter estimation by continuous probing \cite{silb05}, where the so-called Fisher Information is, in fact, given by an ensemble average of products of elements of density matrices that solve stochastic master equations \cite{gamm13_bay}.

\begin{acknowledgements}
We acknowledge A. N. Jordan and A. Jadbabaie for helpful comments.  K.W.M acknowledges support from the John Templeton Foundation and the Sloan Foundation.  K.M. acknowledges support from the Villum Foundation.
\end{acknowledgements}

%



*murch@physics.wustl.edu




\pagebreak

\begin{widetext}

\section{EXPERIMENTAL METHODS AND MEASUREMENT CALIBRATION}

Figure 4 shows a detailed diagram of the microwave setup for the experiment which uses the same qubit and amplifier circuits as described in previous work \cite{tan15}.

\begin{figure}\begin{center}
\includegraphics[angle = 0, width =.8\textwidth]{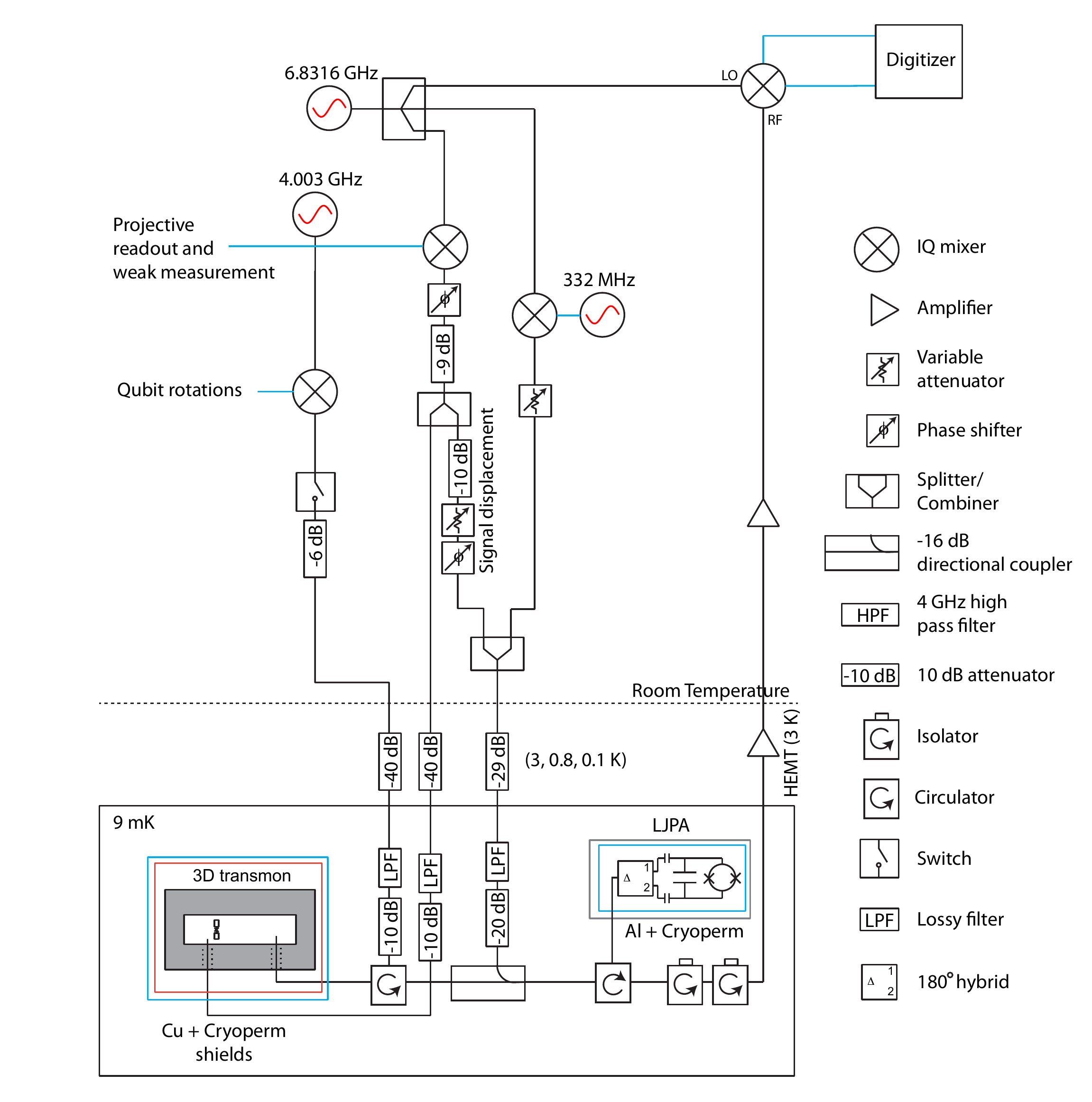}
\end{center}
\caption{ \label{scheme} Experimental setup. The weak measurement is always on except when the projective measurements are performed.}
\end{figure}

To calibrate the parameters of the weak continuous measurement we perform separate calibration experiments.  We prepare the qubit in the $+z$ or $-z$ states using a projective herald measurement and then digitize the ensuing weak measurement.   These weak measurement results are Gaussian distributed and we scale the recorded signal such that the distributions are centered at $+1$ and $-1$ for the $+z$ and $-z$ states respectively, $P(V_t|+z)=(2 \pi a^2)^{-1/2} \exp [-\frac{(V_t-1)^2}{2a^2}]$, $P(V_t|-z)=(2 \pi a^2)^{-1/2} \exp [-\frac{(V_t+1)^2}{2a^2}]$.    The quantum measurement efficiency $\eta = \eta_\mathrm{col}\eta_\mathrm{amp}$ depends on both the collection efficiency and added noise from the amplifiers and is determined from the variance, $a^2 = 1/4 k \eta \Delta t = \kappa/(16 \chi^2 \bar{n} \eta \Delta t)$ where $k=4\chi^2 \bar{n}/\kappa$ represents the measurement strength and $\Delta t$ is the integration time. The parameter $k$ is also related to the characteristic measurement time $\tau$ \cite{webe14} as $k = 1/4\eta\tau$. We used Ramsey measurements to calibrate both the ac Stark shift $(2\chi \bar{n})$ and the dephasing rate $(\Gamma_{m}=8\chi^2\bar{n}/\kappa)$ for photon numbers ranging between $\bar{n}=0$ and $\bar{n}=1.3$, thus determining $\bar{n}$ and $\chi$.

\section{State propagation}

 We propagate $\rho$ forward from the initial state $\rho^{00}=0.95$, $\rho^{01}=0$ when the initial qubit state is $+z$ and $\rho^{00}=0.05$, $\rho^{01}=0$ when the initial state is  $-z$.  
 The stochastic master equation displayed in the main text omits qubit state dephasing for clarity, but the quantum trajectories are calculated including this dephasing which is characterized by the  rate $\gamma=1/T^\ast_2$, where $T^\ast_2 = 16 \ \mu$s. To account for this, $k$ should be replaced by $k+\gamma/2$ in the second term in Eq.\ (3), and in Eq.\ (4) and (9) in the main text. The full stochastic master equation is given by, 
 \begin{eqnarray}
\frac{d \rho}{dt} = -\frac{i}{\hbar} \frac{\Omega_R}{2} [\sigma_y,\rho]  + (k+\frac{\gamma}{2}) (\sigma_z\rho\sigma_z - \rho)+ 2\eta k(\sigma_z\rho+\rho \sigma_z - 2 \mathrm{Tr}(\sigma_z \rho)\rho )V_t. \label{eq:rho}
\end{eqnarray}
 The Rabi frequency  is $\Omega_R/2\pi = 1.16$ MHz and we use time steps of $\Delta t=20$ ns. We perform quantum state tomography on an ensemble of experimental iterations with similar measurement values to verify that we have accurately inferred the quantum trajectory as we have done in previous work \cite{murc13traj,webe14,tan15}. 


\section{Analytical expressions for  $\overline{V}_{WP}$}

From Eq.\ (8) in the main text we can calculate the conditioned expectation value of the voltage signal at time $t$ \cite{tan15}.  The joint probability of measurement outcome  $V_t$ and subsequent measurement outcomes and evolution accumulated in the matrix $E$ can be evaluated, 
\begin{eqnarray}
P_p(V_t) \propto \rho^{00} E^{00} e^{-(V_t-1)^2/2a^2}+\rho^{11} E^{11} e^{-(V_t+1)^2/2a^2}+(\rho^{10}E^{01}+\rho^{01}E^{10})e^{-(V_t^2+1)/2a^2}.
\end{eqnarray}
Where the subscript $p$ denotes ``past" in that the probability depends on both earlier and later probing results. Thus the mean value is given by $\overline{V}_p = \int P_p(V) V dV$ and can be evaluated. In the limit of weak measurement ($a\rightarrow \infty$) we have, 
\begin{equation}\label{condmean}
\overline{V}_{PS} = \frac{\rho^{00}E^{00}-\rho^{11}E^{11}}
{\rho^{00}E^{00}+\rho^{11}E^{11}+\rho^{01}E^{10}+\rho^{10}E^{01}},
\end{equation}
where the elements of $\rho$ and $E$ are all evaluated at time $t$.

The deterministic master equations for $\rho$ and $E$ (Eq.\ (4) and (9) in the main text) can be solved analytically. The two components of $\rho$ are given by
\begin{equation}
\rho^{00}(t)=e^{-kt}(\frac{k}{2\Gamma})[\sin(\Gamma t)+\frac{\Gamma}{k}\cos(\Gamma t)]+\frac{1}{2},
\end{equation} 
\begin{equation}
\rho^{01}(t)=e^{-kt}(\frac{\Omega_R}{2\Gamma})\sin(\Gamma t),
\end{equation} 
where $\Gamma^2 =\Omega_R^2-k^2$ and assuming the qubit is in the ground state at $t=0$.
The analytical solution for the backward evolution, Eq. (9), can be obtained as well for the final state $+z$ at $t=T$.
\begin{equation}
E^{00}(t)=e^{-k(T-t)}(\frac{k}{2\Gamma})[\sin(\Gamma (T-t))+\frac{\Gamma}{k}\cos(\Gamma (T-t))]+\frac{1}{2},
\end{equation} 
\begin{equation}
E^{01}(t)=-e^{-k(T-t)}(\frac{\Omega_R}{2\Gamma})\sin(\Gamma (T-t)).
\end{equation}

\section{Comparison of pre-selected and post-selected measurement signals.}

Figure 5 shows the averaged measurement signals conditioned on the herald measurement and the final projective measurement.  Aside from a slight difference at the beginning (end) which arises from transients in the measurement strength resulting from the herald measurement, the two curves are in close agreement.

\begin{figure}\begin{center}
\includegraphics[angle = 0, width =.5\textwidth]{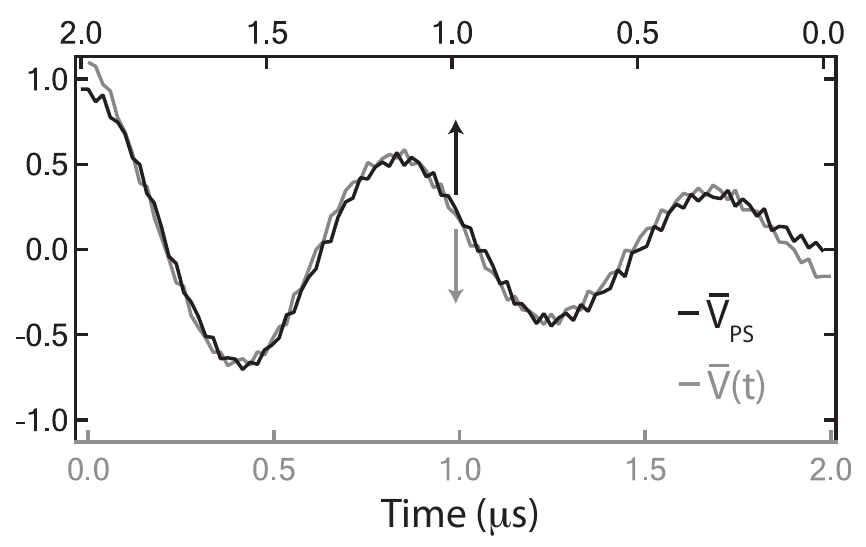}
\end{center}
\caption{ \label{scheme} Pre-selection and Post-selection average measurement signals. The black curve (top axis) corresponds to the average measurement signals from the qubit post-selected in the ground state, which is plotted in reverse.  The gray curve (bottom axis) is the average measurement signal corresponding to pre-selection in the ground state.}
\end{figure}

\end{widetext}

\end {document}